\documentclass[final,5p,times,twocolumn]{elsarticle}

\usepackage{graphicx}
\usepackage{amssymb}
\usepackage{lineno}
\usepackage{rotating}

\journal{NIM A}

\newcommand{\etal}{{\em et al.}}
\newcommand{\PR}{{\em Phys. Rev. }}
\newcommand{\PRL}{{\em Phys. Rev. Lett. }}

\newcommand{\NIM}{{\em Nucl. Instr. Meth. }}

\begin{document}

\begin{frontmatter}

\title{An Electromagnetic Calorimeter for the JLab 
Real Compton Scattering Experiment} 

\author[GLU]{D.~J.~Hamilton\corref{cor}}
\cortext[cor]{Tel.: +44-141-330-5898; Fax: +44-141-330-5889}
\ead{d.hamilton@physics.gla.ac.uk}
\author[YEREVAN]{A. Shahinyan}
\author[JLAB]{B.~Wojtsekhowski}
\author[GLU]{J.~R.~M.~Annand}
\author[UIUC]{T.-H.~Chang}
\author[JLAB]{E.~Chudakov}
\author[UIUC]{A.~Danagoulian}
\author[JLAB]{P.~Degtyarenko}
\author[YEREVAN]{K.~Egiyan\fnref{now}}
\author[RUTGERS]{R.~Gilman}
\author[KHAR]{V.~Gorbenko}\author[GEOR]{J.~Hines\fnref{now1}}
\author[YEREVAN]{E.~Hovhannisyan\fnref{now}}
\author[ODU]{C.~E.~Hyde-Wright}
\author[JLAB]{C.W.~de Jager}
\author[YEREVAN]{A.~Ketikyan}
\author[YEREVAN,JLAB]{V.~H.~Mamyan\fnref{now2}}
\author[JLAB]{R.~Michaels}
\author[UIUC]{A.~M.~Nathan}
\author[SPNPI]{V.~Nelyubin}
\author[BINP]{I.~Rachek}
\author[UIUC]{M.~Roedelbrom}
\author[YEREVAN]{A.~Petrosyan\fnref{now}}
\author[KHAR]{R.~Pomatsalyuk}
\author[JLAB]{V.~Popov}
\author[JLAB]{J.~Segal}
\author[BINP]{Y.~Shestakov}
\author[GEOR]{J.~Templon\fnref{now3}}
\author[YEREVAN]{H.~Voskanyan}

\fntext[now]{deceased}
\fntext[now1]{present address: Applied Biosystems/MDS, USA}
\fntext[now2]{present address: University of Virginia, Charlottesville, VA 22901, USA}
\fntext[now3]{present address: NIKHEF, 1009 DB Amsterdam, The Netherlands}

\address[GLU]{\mbox{University of Glasgow, Glasgow G12 8QQ, Scotland, UK}}
\address[YEREVAN]{\mbox{Yerevan Physics Institute, Yerevan 375036, Armenia}}  
\address[JLAB]{\mbox{Thomas Jefferson National Accelerator Facility,~
Newport News, VA 23606, USA}}
\address[UIUC]{\mbox{University of Illinois, Urbana-Champaign, IL 61801, USA}}
\address[RUTGERS]{\mbox{Rutgers University, Piscataway, NJ 08855, USA}}
\address[KHAR]{\mbox{Kharkov Institute of Physics and Technology,~
Kharkov 61108, Ukraine}}
\address[GEOR]{The University of Georgia, Athens, GA 30602, USA}
\address[ODU]{\mbox{Old Dominion University, Norfolk, VA 23529, USA}}
\address[SPNPI]{\mbox{ St.~Petersburg Nuclear Physics Institute,~
Gatchina, 188350, Russia}}
\address[BINP]{\mbox{Budker Institute for Nuclear Physics,~
Novosibirsk 630090, Russia}}

\begin{abstract}
A lead-glass hodoscope calorimeter that was constructed for use in the Jefferson Lab Real Compton Scattering 
experiment is described. The detector provides a measurement of the coordinates and the energy of scattered photons in the 
GeV energy range with resolutions of 5 mm and 6\%/$\sqrt{E_\gamma \, [GeV]}$.
Features of both the detector design and its performance in the high luminosity environment during the experiment are presented.
\end{abstract}

\begin{keyword}
Calorimeters \sep {\v C}erenkov detectors 
\PACS  29.40Vj \sep 29.40.Ka
`\end{keyword}

\end{frontmatter}

\section{Introduction}
\label{sec:introduction}

\begin{figure}[ht]
 \begin{center}
    \includegraphics[angle=0,width=0.65\linewidth]{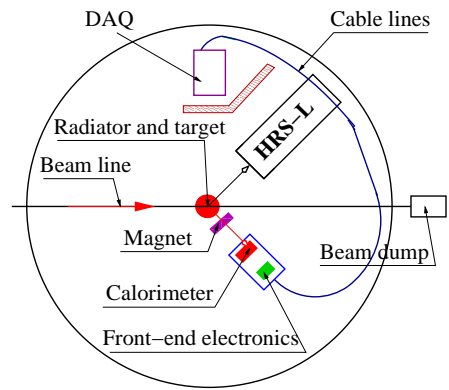}
    \caption{Layout of the RCS experiment in Hall A.
An electron beam incident on a radiator produces an intense 
flux of high energy photons. 
}
    \label{fig:layout}
 \end{center}
\end{figure}

A calorimeter was constructed as part of the instrumentation of
the Jefferson Lab (JLab) Hall A experiment E99-114,
``Exclusive Compton Scattering on the Proton''~\cite{hy99}, 
the schematic layout for which is shown in Fig.~\ref{fig:layout}.
The study of elastic photon scattering 
provides important information about nucleon structure, 
which is complementary to that obtained from elastic electron scattering 
\cite{hy04}.
Experimental data on the Real Compton Scattering (RCS) process at 
large photon energies and large scattering angles are rather scarce, 
due mainly to the absence of high luminosity facilities with suitable 
high-resolution photon detectors. 
Such data are however crucial, as the basic mechanism of the RCS 
reaction is the subject of active debate~\cite{ra98,hu02,to06}.
The only data available before the JLab E99-114 experiment were 
obtained at Cornell about 30 years ago~\cite{sh79}.
\begin{figure*} [tbh]
 \begin{center}
    \includegraphics[angle=0,width=0.8\linewidth]{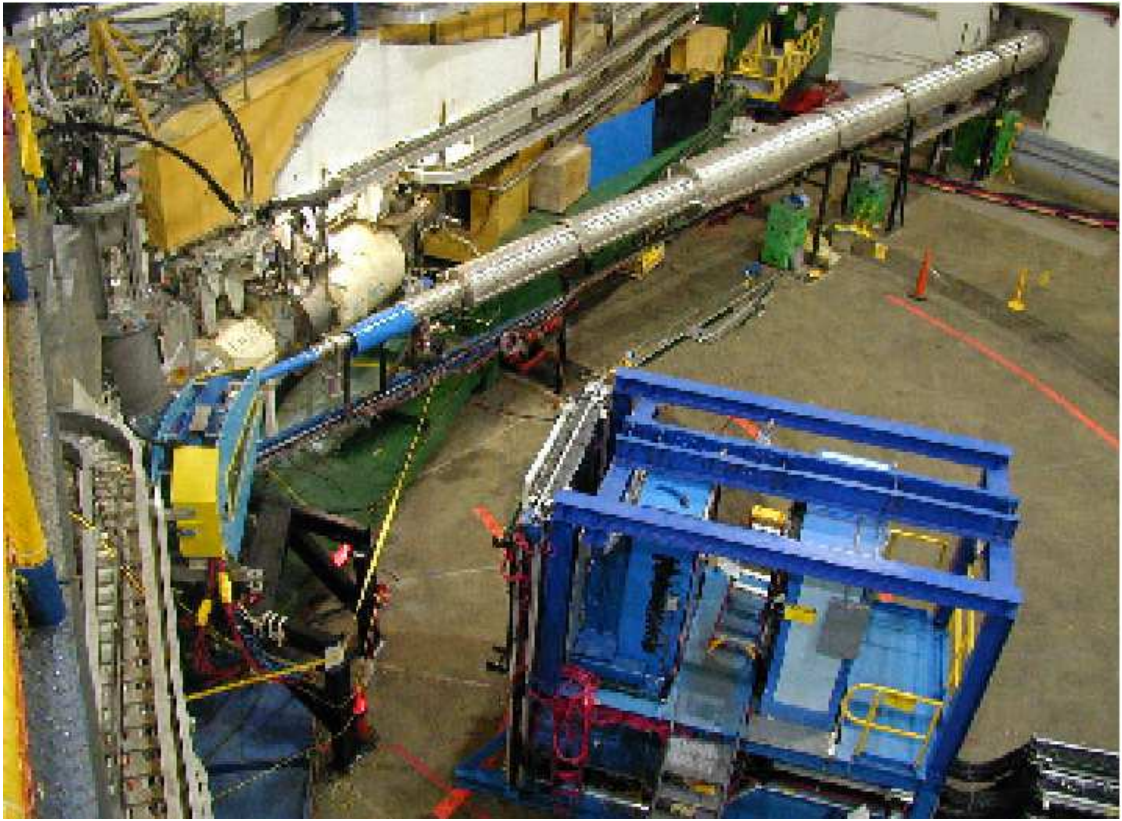}
    \caption{   A photograph of the experimental set-up for E99-114, showing
		the calorimeter (center) and part of the proton spectrometer (rear). }
    \label{fig:picture}
 \end{center}
\end{figure*}

The construction of the CEBAF (Continuous Electron Beam Accelerator Facility) 
accelerator has led to an extension of 
many experiments with electron and photon beams in the GeV energy range 
and much improved precision. 
This is the result of a number of fundamental improvements to the electron
beam, including a 100\% duty cycle, low emittance and 
high polarization, in addition to new dedicated target and detector systems. 
The CEBAF duty factor provides an improvement of a factor of 15 compared 
to the best duty factor of a beam extracted from a synchrotron, at a similar
instantaneous rate in the detectors.

In 1994 work began on the development of a technique for an RCS experiment at JLab,  
leading in 1997 to the instigation of a large-scale prototyping effort.
The results of the subsequent test runs in 1998 and 1999~\cite{ch98} provided sufficient 
information for the final design of the apparatus presented in the present article.
The fully realized physics experiment took place in 2002 (see Fig.~\ref{fig:layout}) at a photon-nucleon 
luminosity which was a factor of 1300 higher than in the previous Cornell experiment.
The experimental technique involves utilizing a mixed electron-photon beam which 
is incident on a liquid hydrogen target
and passes to a beam dump. The scattered photons are detected in the calorimeter, while the
recoiling protons are detected in a high resolution magnetic spectrometer (HRS-L).
A magnet between the hydrogen target and the calorimeter deflects the
scattered electrons, which then allows for clean separation between 
Compton scattering and elastic e-p scattering events.
The Data Acquisition Electronics (DAQ) is shielded by a 4 inch thick concrete
wall from the beam dump and the target.
Figure~\ref{fig:picture} shows a photograph of the experimental set-up with
the calorimeter in the center.

The experiment relied on
a proton-photon time coincidence and an accurate measurement 
of the proton-photon kinematic correlation for event selection.
The improvement in the event rate over the previous measurement was achieved 
through the use of a mixed electron-photon beam, which in turn required 
a veto detector in front of the calorimeter or the magnetic deflection 
of the scattered electron \cite{hy99}. In order to ensure redundancy and cross-checking, 
both a veto and deflection magnet were designed and built.
The fact that a clean photon beam was not required meant that
the photon radiator could be situated very close to the hydrogen target,  
leading to a much reduced background near the beam line and a 
dramatic reduction of the photon beam size.
This small beam size in combination with the large dispersion in 
the HRS-L proton detector system \cite{al04} resulted in very good momentum and angle 
resolution for the recoiling proton without the need for a tracking detector 
near the target, where the background rate is high. 

Good energy and coordinate resolutions were key features of the photon
detector design goals, both of which were significantly improved in the 
JLab experiment as compared to the Cornell one.
An energy resolution of at least 10\% is required to separate cleanly 
RCS events from electron bremsstrahlung and neutral pion events.
In order to separate further the background from neural pion photo-production, which is the 
dominant component of the high-energy background in this measurement, 
a high angular resolution between proton and photon detectors is crucial. 
This was achieved on the photon side by constructing a highly 
segmented calorimeter of 704 channels.
The RCS experiment was the first instance of a calorimeter being operated 
at an effective electron-nucleon luminosity of $10^{39}$~cm$^2$/s 
\cite{ha05, da07} (a 40~$\mu$A electron beam on a 6\% Cu 
radiator upstream of a 15~cm long liquid hydrogen target).
It was observed in the test runs that the counting rate in the calorimeter 
fell rapidly as the threshold level was increased, which presented an opportunity 
to maintain a relatively low trigger rate even at high luminosity.
However, on-line use of the calorimeter signal required a set of summing electronics 
and careful equalizing and monitoring of the individual channel outputs during the experiment.

As the RCS experiment represented the first use of such a calorimeter at very high luminosity, 
a detailed study of the calorimeter performance throughout the course of the experiment has been 
conducted. This includes a study of the relationship between luminosity, trigger rate, energy resolution 
and ADC pedestal widths. An observed fall-off in energy resolution as the experiment progressed allowed 
for characterization of radiation damage sustained by the lead-glass blocks. It was possible to mitigate
this radiation damage after the experiment by annealing, with both UV curing and heating proving effective.

We begin by discussing the various components which make up the calorimeter and the methods used in their
construction. This is followed by a description of veto hodoscopes which were used for particle identification 
purposes. An overview of the high-voltage and data acquisition systems is then presented, followed, finally, by 
a discussion on the performance of the calorimeter in the unique high-luminosity environment during the RCS
experiment.

\section{Calorimeter}
\label{sec:calorimeter}
\begin{figure} [ht]
 \begin{center}
    \includegraphics[angle=0,width=\linewidth]{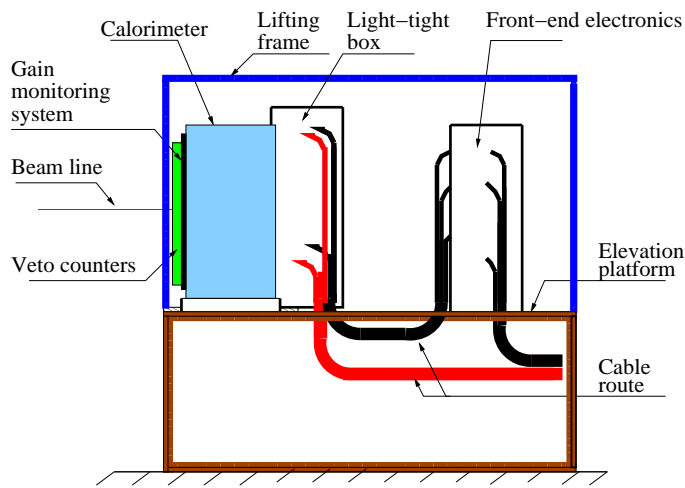}
     \caption{Schematic side view of the RCS calorimeter detector system.}
    \label{fig:side-view}
  \end{center}
\end{figure}

The concepts and technology associated with a fine-granularity 
lead-glass {\v C}erenkov electromagnetic calorimeter (GAMS) were developed 
by Yu.~Prokoshkin and collaborators at the Institute of High Energy 
Physics (IHEP) in Serpukhov, Russia \cite{pr86}.
The GAMS type concept has since been employed for detection
of high-energy electrons and photons in 
several experiments at JLab, IHEP, CERN, 
FNAL and DESY (see for example \cite{selex}).
Many of the design features of the calorimeter presented in this article 
are similar to those of Serpukhov. A schematic showing the overall design of 
the RCS calorimeter can be seen in Fig.~\ref{fig:side-view}. 
The main components are:
\begin{itemize}
\item the lead-glass blocks;
\item a light-tight box containing the PhotoMultiplier Tubes (PMTs);
\item a gain-monitoring system;
\item a doubly-segmented veto hodoscopes;
\item the front-end electronics;
\item an elevated platform;
\item a lifting frame. 
\end{itemize}

The calorimeter frame hosts a matrix of 22$\times$32 lead-glass blocks
together with their associated PMTs and High Voltage (HV) dividers.
Immediately in front of the lead-glass blocks is a sheet of UltraViolet-Transmitting (UVT) Lucite,
which is used to distribute calibration light pulses
for gain-monitoring purposes uniformly among all 704 blocks.
The light-tight box provides protection of the PMTs from ambient 
light and contains an air-cooling system as well as the HV 
and signal cable systems.
Two veto hodoscopes, operating as {\v C}erenkov counters with UVT Lucite 
as a radiator, are located in front of the calorimeter. 
The front-end electronics located a few feet behind the detector 
were assembled in three relay racks.
They are comprised of 38 analog summers, trigger logic and patch panels.
The elevated platform was needed to bring the calorimeter to the level of
the beam line, while the lifting frame was used to re-position the calorimeter 
in the experimental hall by means of an overhead crane.
This procedure, which took on average around two hours, 
was performed more than 25 times during the course of the experiment.

\subsection{Calorimeter Design}
\begin{figure}[ht]
 \begin{center}
   \includegraphics[angle=0,width=\linewidth]{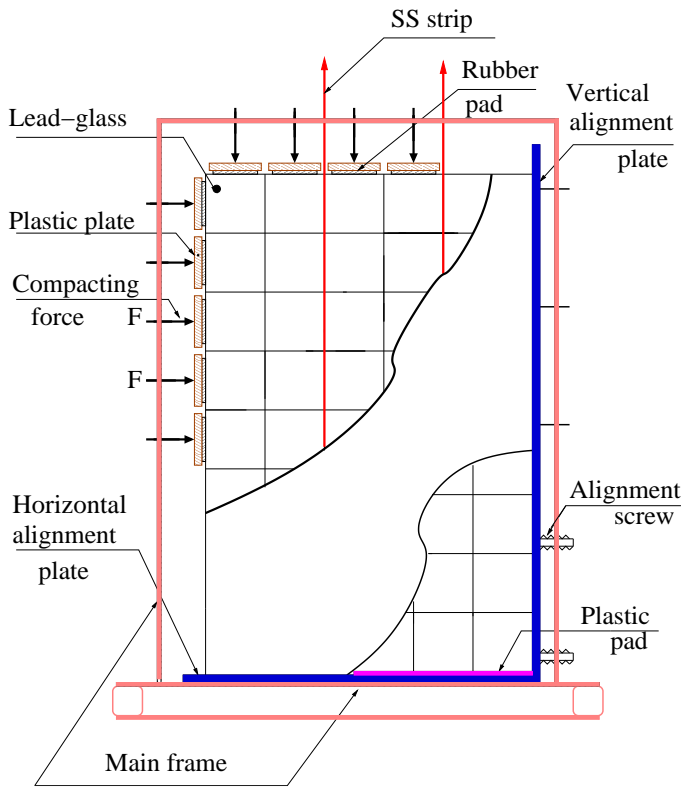}
   \caption{Front cross-section of the calorimeter, showing the mechanical components.}
   \label{fig:front}
 \end{center}
\end{figure}

%
The main frame of the calorimeter is made of 10 inch wide steel C-channels.
A thick flat aluminum plate was bolted to the bottom of the frame, with a 
second plate installed vertically and aligned to 90$^\circ$
with respect to the first one by means of alignment screws 
(see Fig.~\ref{fig:front}).
Another set of screws, mounted inside and at the top of the main frame 
on the opposite side of the vertical alignment plate, was used to compress 
all gaps between the lead-glass modules and to fix their positions. 
The load was applied to the lead-glass blocks through 
1~inch~$\times$~1~inch~$~\times$~0.5~inch plastic plates and a 0.125 inch 
rubber pad.
In order to further assist block alignment, 1 inch wide stainless steel 
strips of 0.004 inch thickness running from top to bottom of the frame 
were inserted between every two columns of the lead-glass modules.

\subsubsection{Air Cooling}
%
\begin{figure}[ht]
 \begin{center}
   \includegraphics[angle=0,width=0.75\linewidth]{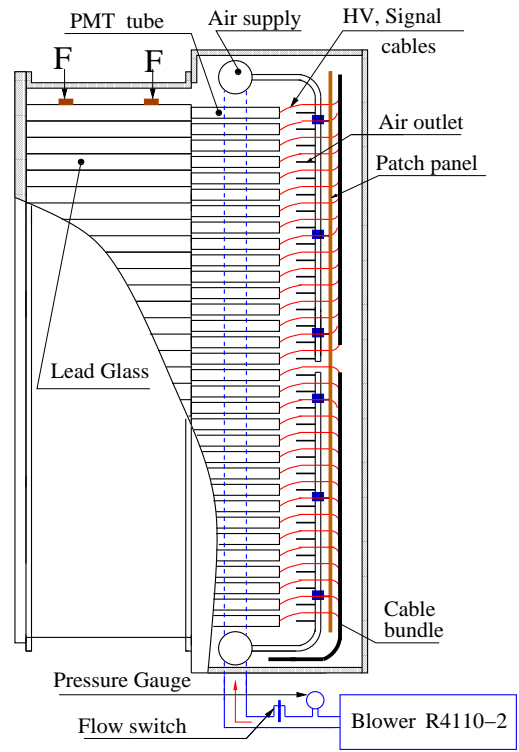}
    \caption{A schematic showing the calorimeter air cooling and cabling systems.}
   \label{fig:cooling}
 \end{center}
\end{figure}
All PMTs and HV dividers are located inside a light-tight box, as 
shown in Fig.~\ref{fig:cooling}. 
As the current on each HV divider is 1~mA, simultaneous 
operation of all PMTs would, without cooling, lead to a temperature 
rise inside the box of around 50-70$^\circ$C. 
An air-cooling system was developed to prevent the PMTs from 
overheating, and to aid the stable operation of the calorimeter.
The air supply was provided by two parallel oil-less regenerative 
blowers of R4110-2 type\footnote{Manufactured by S\&F Supplies, Brooklyn, NY 11205, USA.}, 
which are capable of supplying
air at a maximum pressure of 52 inches water and 
a maximum flow of 92~CFM.
The air is directed toward the HV divider via vertical collector 
tubes and numerous outlets. When the value on any one of the temperature 
sensors installed in several positions inside the box exceeds a preset
limit, the HV on the PMTs is turned off by an interlock system.    
The air line is equipped with a flow switch of type FST-321-SPDT 
which was included in the interlock system. The average temperature 
inside the box during the entire experimental run did not exceed the 
preset limit of 55$^{\circ}$C.

\subsubsection{Cabling System}
\begin{figure}[htb] 
 \begin{center}
   \includegraphics[angle=0,width=\linewidth]{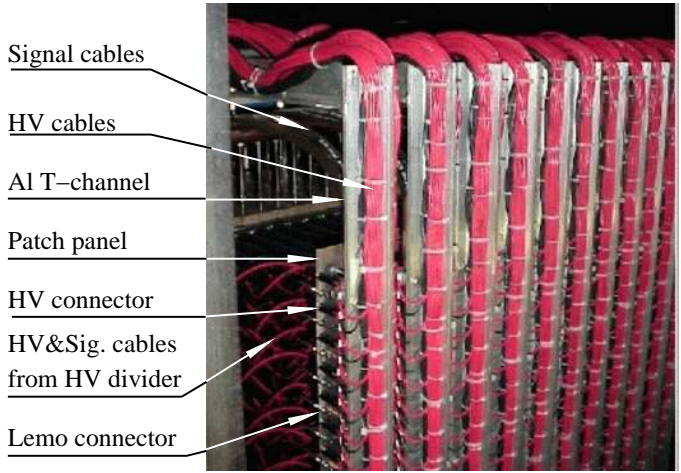}
    \caption{A photograph of the cabling inside the PMT box.}
   \label{fig:cabling}
 \end{center}
\end{figure}

A simple and reliable cabling system is one of the key 
features of multichannel detectors, with easy access 
to the PMTs and HV dividers for installation and repair being one
of the key features.
The cabling system includes:
\begin{itemize}
\item 1 foot long HV and signal pig-tails soldered to the HV divider;
\item patch panels for Lemo and HV connectors;
\item 10 feet long cables from those patch panels to the 
front-end electronics and the HV distribution boxes;
\item the HV distribution boxes themselves;
\item BNC-BNC patch panels for the outputs of the front-end modules;
\item BNC-BNC patch panels on the DAQ side for the analog signals;
\item BNC-Lemo patch panels on the DAQ side for the veto-counter lines.
\end{itemize}
Figure \ref{fig:cabling} shows the cabling arrangement inside the PMT box.
The patch panels, which are custom-built and mounted on the air supply 
tubes, have the ability to swing to the side in order to allow access 
to the PMTs and the HV dividers. The box has two moving doors, the opening 
of which leads to activation of an interlock system connected to the HV supply.
In order to reduce the diameter of the cable bundle from the PMT box, 
RG-174 cable (diameter 0.1 inch) was used for the PMT signals,
and a twisted  pair for the HV connection (two individually insulated
inner 26 AWG conductors with an overall flame-retardant PVC 
jacket, part number 001-21803 from the General Wire Product company).
The box patch panels used for the HV lines each convert 24 of the above 
twisted pairs (single HV line) to the multi-wire HV cables 
(the part 001-21798 made by General Wire Product),
which run to the HV power supply units located in the shielded
area near DAQ racks.

\subsection{Lead-Glass Counter}
The basic components of the segmented calorimeter are 
the TF-1 lead-glass blocks and the FEU 84-3 PMTs.
In the 1980s the Yerevan Physics Institute (YerPhI) purchased a 
consignment of TF-1 lead-glass blocks of 4~cm~$\times~$4~cm$~\times$40~cm 
and FEU~84-3 PMTs of 34~mm diameter (with an active photo-cathode diameter 
of 25~mm) for the construction of a calorimeter 
to be used in several experiments at the YerPhI synchrotron.
In January of 1998 the RCS experiment at JLab was approved
and soon after these calorimeter components were shipped from 
Yerevan to JLab. This represented the YerPhi contribution to the experiment,
as the properties of the TF-1 lead-glass met the requirements of the experiment 
in terms of photon/electron detection with reasonable energy and position resolution 
and radiation hardness. The properties of TF-1 lead-glass \cite{selex, ast73} are given in Table 
\ref{tab:tf1}.

\begin{table}[th]
\caption{Important properties of TF-1 lead-glass.}
\vspace{ 0.1 in}
\begin{center}
{\footnotesize
\begin{tabular}{|l|c|}
\hline
Density             &  $3.86~gcm^{-3}$ \\ 
\hline
Refractive Index    &  $1.65$        \\ 
\hline
Radiation Length    &  $2.5~cm$      \\ 
\hline
Moli{\'e}re Radius  &  $3.50~cm$      \\ 
\hline
Critical Energy     &  $15~MeV$      \\ 
\hline
\end{tabular}}
\end{center}
\label{tab:tf1}   
\end{table}

All PMTs had to pass a performance test with
the following selection criteria: 
a dark current less than 30~nA, a gain of 10$^6$ with
stable operation over the course of the experiment (2 months), a linear dependence 
of the PMT response (within 2 \%) on an incident optical pulse of 300 to 30000 photons. 
704 PMTs out of the 900 available were selected as a result of these
performance tests. Furthermore, the dimensional tolerances were checked for all lead-glass blocks, 
with strict requirements demanded on the length (400$\pm$2~mm) and 
transverse dimensions (40$\pm$0.2~mm).  
\begin{figure*}[tbh] 
  \begin{center}
    \includegraphics*[width= 0.75 \linewidth]{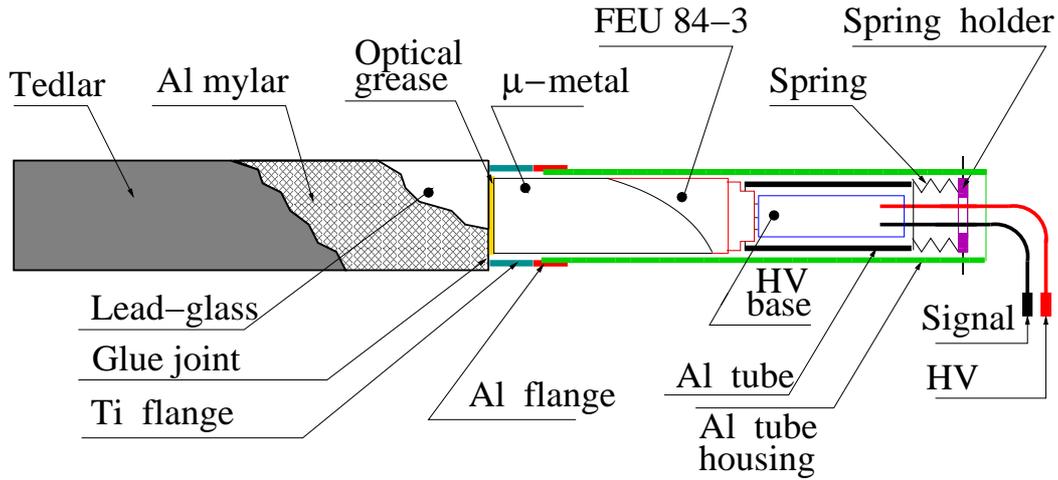}
       \caption{Schematic of the lead-glass module structure.}
    \label{fig:counter}
  \end{center}
\end{figure*}
\subsubsection{Design of the Counter}
In designing the individual counters for the RCS calorimeter, 
much attention was paid to reliability, simplicity and the
possibility to quickly replace a PMT and/or HV divider.
The individual counter design is shown in Fig.~\ref{fig:counter}.
A titanium flange is glued to one end of the lead-glass 
block by means of EPOXY-190. Titanium was selected 
because its thermal expansion coefficient is very close 
to that of the lead glass. The PMT housing, which is bolted to 
the Ti flange, is made of an anodized Al flange and an Al tube. 
The housing contains the PMT and a $\mu$-metal shield, 
the HV divider, a spring, a smaller Al tube which transfers
a force from the spring to the PMT, and a ring-shaped spring holder.
The optical contact between the PMT and the lead-glass block
is achieved by use of optical grease, type BC-630 (Bicron),
which was found to increase the amount of light detected by 
the PMT by 30-40\% compared to the case without grease. The PMT is pressed to the lead-glass block by means of 
a spring, which pushes the HV base with a force of 0.5-1~lbs.
Such a large force is essential for the stability of 
the optical contact over time at the elevated temperature of the PMTs.
The glue-joint between the lead glass and the Ti flange, which
holds that force, failed after several months in a significant 
fraction (up to 5\%) of the counters. 
An alternative scheme of force compensation was realized 
in which the force was applied to the PMT housing from the external 
bars placed horizontally between the PMT housing 
and the patch-panel assembly.
Each individual lead-glass block was wrapped in aluminized Mylar film 
and black Tedlar (a polyvinyl fluoride film from DuPont) for optimal light collection and inter-block isolation.
Single-side aluminized Mylar film was used with the Al layer 
on the opposite side of the glass. 
Such an orientation of the film limits the diffusion of 
Al atoms into the glass and the non-oxidized surface 
of aluminum, which is protected by Mylar, provides a better reflectivity.
The wrapping covers the side surface of the lead-glass block, 
leaving the front face open for the gain monitoring. 
The signal and the HV cables are each one foot long.
They are soldered to the HV divider on one end and terminated with 
Lemo\copyright00 and circular plastic connectors (cable mount receptacle from Hypertronics) on the other end.
The cables leave the PMT housing through the open center of the spring holder.
\subsubsection{HV Divider}
\begin{figure}[ht] 
  \begin{center}
    \includegraphics[angle=0,width=\linewidth]{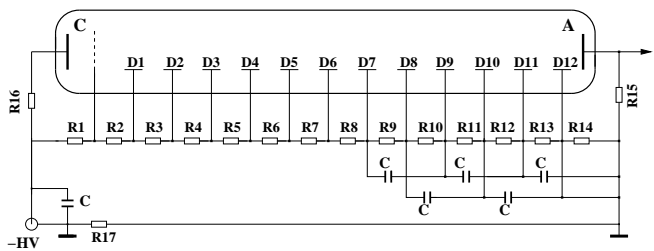}
    \caption{Schematic of the high-voltage divider for the FEU~84-3 PMT.
      The values of the resistors are $R(1-10)=100~k\Omega, R11=130~k\Omega, 
      R12=150~k\Omega, R13=200~k\Omega, R14=150~k\Omega, R15=10~k\Omega, 
      R16=10~M\Omega, R17=4~k\Omega$. The capacitance C is 10~nF. 
    }
    \label{fig:divider}
  \end{center}
\end{figure}

%
At the full luminosity of the RCS experiment ($0.5 \times 10^{39}$~cm$^2$/s) 
and at a distance of 6~m from the target the background energy load 
per lead-glass block reaches a level of 10$^8$~MeVee (electron equivalent) 
per second, which was found from the average value of anode current in
the PMTs and the shift of the ADC pedestals for a 150 ns gate width.
At least 30\% of this energy flux is due to high energy particles
which define the counting rate.
The average energy of the signals for that component,
according to the observed rate distribution, is in the range of 
100-300~MeVee, depending on the beam energy and the detector angle.
The corresponding charge in the PMT pulse is around 5-15~pC collected in 10-20~ns. 
The electronic scheme and the selected scale of 1~MeV per ADC channel (50~fC) 
resulted in an average anode current of 5~$\mu$A due to background load.
A high-current HV base (1~mA) was therefore chosen to reduce the effect 
of the beam intensity variation on the PMT amplitude and 
the corresponding energy resolution to the level of 1\%.
The scheme of the HV base is shown in Fig.~\ref{fig:divider}.
According to the specification data for the FEU~84-3 PMTs
the maximum operation voltage is 1900~V. 
Therefore a nominal voltage value of 1800~V 
and a current value in the voltage divider of 1~mA were chosen.
\subsection{Electronics}
\label{sec:electronics}

\begin{figure*}[bht]
  \begin{center}
    \includegraphics[angle=0,width=\linewidth]{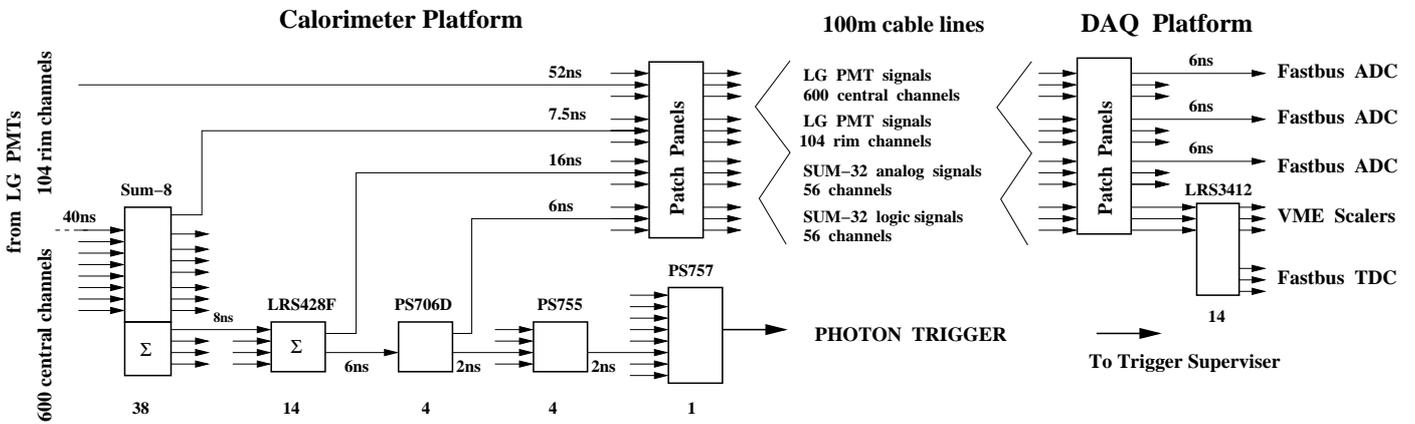}
   \caption{A block diagram of the calorimeter electronics.}
   \label{fig:block-diagram}
  \end{center}
\end{figure*}

The calorimeter electronics were distributed over two locations; see
the block diagram in Fig.~\ref{fig:block-diagram}.
The first group of modules (front-end) is located in three racks mounted 
on the calorimeter platform in close vicinity to the lead-glass blocks. 
These are the trigger electronics modules which included a mix of 
custom-built and commercially available NIM units:
\begin{itemize}
\item 38 custom-built analog summing modules used for level-one signal summing 
\footnote{This module was designed by S. Sherman, Rutgers University.};
\item 14 linear fan-in/fan-out modules (LeCroy model 428F) for
a second-level signal summation;
\item 4 discriminator units (Phillips Scientific model 706);
\item a master OR circuit, realized with Phillips Scientific logic units 
(four model 755 and one model 757 modules);
\item several additional NIM modules used to provide 
auxiliary trigger signals for the calorimeter calibration with cosmics and for the
PMT gain-monitoring system.
\end{itemize}

The second group of electronic modules, which include charge and time 
digitizers as well as equipment for the Data Acquisition, High Voltage 
supply and slow-control systems, is placed behind a radiation-protecting
concrete wall.
All 704 lead-glass PMT signals and 56 {\em SUM-32} signals are digitized 
by LeCroy 1881M FastBus ADC modules. 
In addition, 56 {\em SUM-32} discriminator 
pulses are directed to scalers and to LeCroy 1877 FastBus TDCs.
Further detailed information about the electronics is 
presented in Section~\ref{sec:DAQ}.

The signals between these locations are transmitted via patch-panels 
and coaxial cables, consisting of a total number of 1040~signal 
and 920~HV lines. 
The length of the signal cables is about 100~m, which serve as 
delay lines allowing the timing of the signals at the ADC inputs to be 
properly set with respect to the ADC gate, formed by the experiment trigger.
The width of the ADC gate (150~ns) was made much wider than the 
duration of PMT pulse in order to accommodate the wider pulses 
caused by propagation in the 500~ns delay RG-58 signal cables.
The cables are placed on a chain of bogies, which permits the calorimeter 
platform to be moved in the experimental hall without disconnecting the cables. 
This helped allow for a quick change of kinematics.
\subsubsection{Trigger Scheme}
The fast on-line photon trigger is based on PMT signals from the
calorimeter counters. 
The principle of its operation is a simple constant-threshold method, 
in which a logic pulse is produced if the energy deposition in the calorimeter 
is above a given magnitude. 
Since the Moli\`{e}re radius of the calorimeter material 
is $R_M\approx 3.5$ cm, 
the transverse size of the electromagnetic shower in 
the calorimeter exceeds the size of a single lead-glass block. 
This enables a good position sensitivity of the device, while 
at the same time making it mandatory for the trigger scheme 
to sum up signals from several adjacent counters to get a signal 
proportional to the energy deposited in the calorimeter. 

\begin{figure}[bh]
  \begin{center}
    \includegraphics[angle=0,width= 0.75 \linewidth]{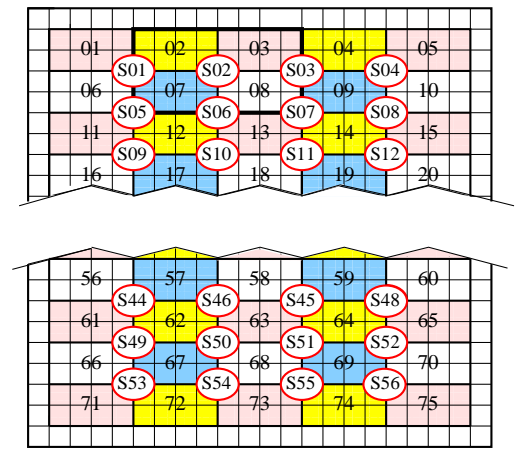}
   \caption{The principle of two-level summation of signals 
for the hardware trigger: 75 eight-block sub-arrays and 56 
overlapping groups of four sub-arrays forming {\em SUM-32} signals
labeled as {\bf S01-S56}.
In the highlighted example the sums 02,03,07, and 08 form a S02 signal.}
    \label{fig:summing}
  \end{center}
\end{figure}

From an electronics point of view, the simplest realization 
of such a trigger would be a summation of {\it all} blocks followed
by a single discriminator. 
However, such a design is inappropriate for a high-luminosity 
experiment due to the very high background level. 
The opposing extreme approach would be to form a summing signal 
for a small group including a single counter hit and its 8 adjacent 
counters, thus forming a \mbox{3 $\times$ 3} block structure.
This would have to be done for {\it every} lead-glass block, 
except for those at the calorimeter's edges, leading to an 
optimal signal-to-background ratio, but an impractical 600 channels of
{\it analog splitter$\rightarrow$analog summer$\rightarrow$discriminator} 
circuitry followed by a 600-input fan-in module.
The trigger scheme that was adopted and is shown in Fig.~\ref{fig:summing} 
is a trade-off between the above extreme cases.
This scheme contains two levels of analog summation followed 
by appropriate discriminators and an OR-circuit.
It involved the following functions: 
\begin{itemize}
\item
the signals from each PMT in the 75 2$\times$4 sub-arrays of
adjacent lead-glass blocks, excluding the outer-most blocks, are
summed in a custom-made analog summing module to 
give a {\em SUM-8} signal (this module duplicates 
the signals from the PMTs with less then 1\% integral nonlinearity);
\item
these signals, in turn, are further summed in overlapping groups of four in
LeCroy LRS428F NIM modules to produce 56 {\em SUM-32} signals. 
Thus, each {\em SUM-32} signal is proportional to the energy deposition 
in a subsection of the calorimeter of 4 blocks high and 8 blocks wide, 
i.e. 16$\times$32 cm$^2$. Although this amounts to only 5\% of the 
calorimeter acceptance, for any photon hit (except for those at the 
edges) there will be at least one segment which contains the whole
electromagnetic shower.
\item
the {\em SUM-32} signals are sent to constant-threshold discriminators,
from which the logical pulses are OR-ed to form the photon singles 
trigger T1 (see Section~\ref{sec:DAQ}).
The discriminator threshold is remotely adjustable, and was 
typically set to around half of the RCS photon energy for a 
given kinematic setting.
\end{itemize}
\subsection{Gain Monitoring System}
\label{sec:monitoring}
The detector is equipped with a system that distributes light pulses to each 
calorimeter module. 
The main purpose of this system is to provide a quick way to check the 
detector operation and to calibrate the dependence of the signal amplitudes on 
the applied HV.
The detector response to photons of a given energy may drift with 
time, due to drifts in the PMT gains and to changes 
in the glass transparency caused by radiation damage. For this reason, 
the gain monitoring system also allowed measurements of 
the relative gains of all detector channels during the experiment. 
In designing the gain-monitoring system ideas developed
for a large lead-glass calorimeter at BNL\cite{Crittenden:1997ug} were used.

\begin{figure}[ht] 
 \begin{center}
   \includegraphics[angle=0,width=0.7\linewidth]{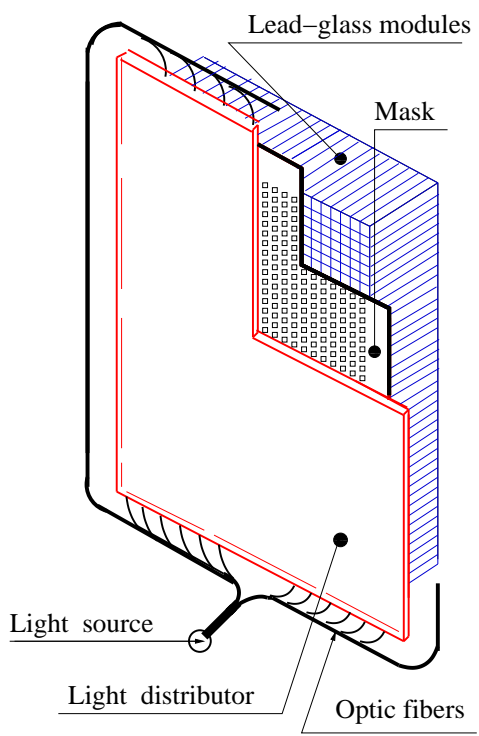}
   \caption{Schematic of the Gain-monitoring system.}
   \label{fig:monitoring}
 \end{center}
\end{figure}

The system includes two components: a stable light source and a 
system to distribute the light to all calorimeter modules. 
The light source consists of an LN300 nitrogen laser\footnote{
Manufactured by Laser Photonics, Inc, FL 32826, USA.}, 
which provides 5~ns long, 300~$\mu$J ultraviolet light pulses 
of 337~nm wavelength. 
The light pulse coming out of the laser is attenuated, typically by
two orders of magnitude, and  monitored using a silicon photo-diode 
S1226-18BQ\footnote{Manufactured by Hamamatsu Photonics, Hamamatsu, Japan.}
mounted at 150$^\circ$ to the laser beam. 
The light passes through an optical filter, several of which 
of varying densities are mounted on a remotely controlled 
wheel with lenses, before arriving at 
a wavelength shifter. 
The wavelength shifter used is a 1 inch diameter semi-spherical 
piece of plastic scintillator, in which the ultraviolet light is fully
absorbed and converted to a blue ($\sim$ 425~nm) light pulse, radiated
isotropically. 
Surrounding the scintillator about 40 plastic fibers (2~mm thick and 4~m 
long) are arranged, in order to transport the light to the sides of a Lucite 
plate. This plate is mounted adjacent to the front face of the 
lead-glass calorimeter and covers its full aperture (see Fig.\ref{fig:monitoring}).
The light passes through the length of the plate, causing it to glow 
due to light scattering in the Lucite. 
Finally, in order to eliminate the cross-talk between adjacent counters
a mask is inserted between the Lucite plate and the detector face. 
This mask, which reduces the cross-talk by at least a factor of 100, 
is built of 12.7~mm thick black plastic and contains a 2~cm~$\times$~2~cm 
hole in front of each module. 

Such a system was found to provide a 
rather uniform light collection for all modules, and proved 
useful for detector testing and tuning, as well as for 
troubleshooting during the experiment. However, it was found that 
monitoring over extended periods of time proved to be less informative 
than first thought. The reason for this is due to the fact that the 
main radiation damage to the lead-glass blocks occurred at a depth of about 
2-4~cm from the front face. The monitoring light passes through the 
damaged area, while an electromagnetic 
shower has its maximum at a depth of about 10~cm. 
Therefore, as a result of this radiation damage the magnitude 
of the monitoring signals drops relatively quicker than the 
real signals. Consequently, the resulting change in light-output during
the experiment was characterized primarily through online analysis of 
dedicated elastic e-p scattering runs. This data was then used for periodic
re-calibration of the individual calorimeter gains.
\section{Veto Hodoscopes}
\label{sec:Veto}
\begin{figure}[htb] 
 \begin{center}
   \includegraphics[angle=0,width=0.75\linewidth]{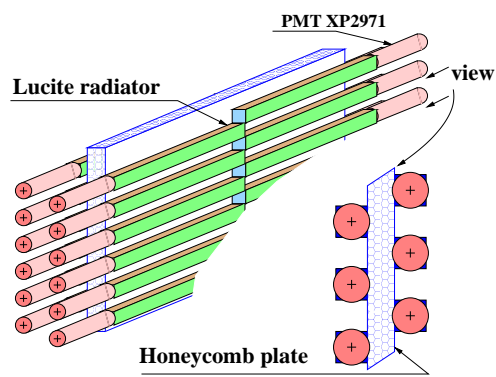}
    \caption{Cut-off view of the ``horizontal'' veto hodoscope.}
   \label{fig:veto-horizontal}
 \end{center}
\end{figure}

In order to ensure clean identification of the scattered photons through
rejection of high-energy electrons in the complicated environment 
created by the mixed electron-photon 
beam, a veto detector which utilizes UVT Lucite as a {\v C}erenkov radiator
was developed. This veto detector proved particularly useful for 
low luminosity runs, where its use
made it possible to take data without relying on the deflection magnet 
(see Fig.~\ref{fig:layout}).
The veto detector consists of two separate hodoscopes located 
in front of the calorimeter's gain monitoring system.
The first hodoscope has 80 counters oriented vertically, while
the second has 110 counters oriented horizontally as shown in
Fig.~\ref{fig:veto-horizontal}.
The segmentation scheme for the veto detector was chosen so that it was 
consistent with the position resolution of the lead-glass calorimeter.
An effective dead time of an individual counter is about 100~ns due 
to combined double-pulse resolution of the PMT, the front-end electronics, 
the TDC, and the ADC gate-width. 

\begin{figure}[htb] 
 \begin{center}
   \includegraphics[angle=0,width=0.75\linewidth]{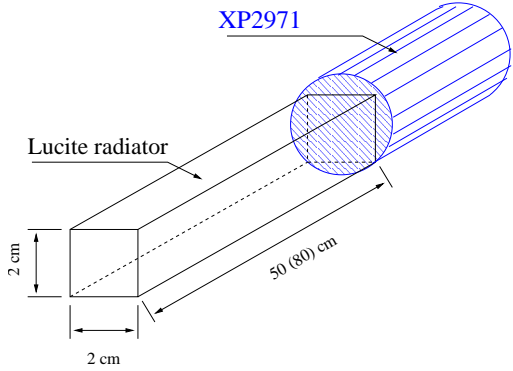}
     \caption{Schematic of the veto counter.}
   \label{fig:veto-counter}
 \end{center}
\end{figure}

Each counter is made of a UVT Lucite bar with a PMT glued directly to 
one of its end, which can be seen in Fig.~\ref{fig:veto-counter}.
The Lucite bar of 2$\times$2~cm$^2$ cross section was glued to a ~XP2971 PMT 
and wrapped in aluminized Mylar and black Tedlar.
Counters are mounted on a light honeycomb plate via an alignment
groove and fixed by tape. The counters are staggered in such a way so 
as to allow for the PMTs and the counters to overlap.

\begin{figure}[ht] 
 \begin{center}
   \includegraphics[angle=0,width=0.9\linewidth]{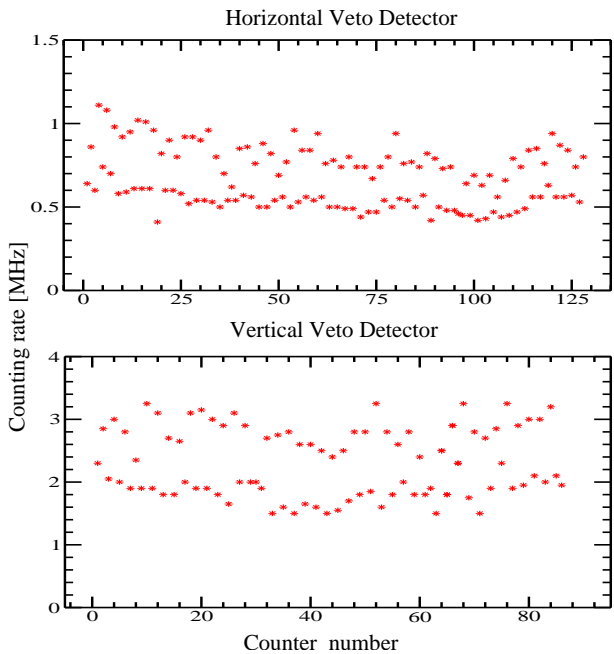}
   \caption{The counting rate in the veto counters observed
at luminosity of $1.5 \cdot 10^{38}$~cm$^{-2}$/s.}
   \label{fig:veto-rates}
 \end{center}
\end{figure}

The average PMT pulse generated by a high-energy electron corresponds to 
20~photo-electrons.
An amplifier, powered by the HV line current, was added to the 
standard HV divider, in order that
the PMT gain could be reduced by a factor of 10 \cite{po01, po03}.
After gain-matching by using cosmic ray data a good rate uniformity 
was achieved, as can be seen in the experimental rate distribution of 
the counters shown in Fig.~\ref{fig:veto-rates}. The regular variation 
in this distribution reflects the shielding effect resulting from 
the staggered arrangement of the counters.
A significant reduction of the rate (by a factor of 5) was achieved
by adding a 2 inch polyethylene plate in front of the hodoscopes.
Such a reduction as a result of this additional shielding is 
consistent with the observed variation of the rate 
(see Fig.~\ref{fig:veto-rates}) and indicates that the typical energy
of the dominant background is around a few MeV.
The veto plane efficiency measured for different beam intensities is shown
in Table~\ref{tab:veto-efficiency}. It drops significantly at high rate
due to electronic dead-time, which limited the beam intensity to 3-5~$\mu$A
in data-taking runs with the veto. 
\begin{table}[th]
\caption{The efficiency of the veto hodoscopes and the rate of a single counter
at different beam currents. The detector was installed at 30$^\circ$ with respect 
to the beam at a distance 13~m from the target. The radiator had been removed
from the beam path, the deflection magnet was off and the 
2 inch thick polyethylene protection plate was installed.  }
\vspace{ 0.1 in}
\begin{center}
{\footnotesize
\begin{tabular}{|l|c|c|c|c|}
\hline
Run     & Beam    &  Rate of the  & Efficiency & Efficiency \\ 
        & current &  counter V12  & horizontal  & vertical   \\
        & [$\mu$A]&   [MHz]       & hodoscope  & hodoscope  \\ 
\hline
1811    &  2.5    &  0.5          &  96.5\%    & 96.8\% \\ 
\hline 
1813    &  5.0    &  1.0          &  95.9\%    & 95.0\% \\ 
\hline
1814    &  7.5    &  1.5          &  95.0\%    & 94.0\% \\ 
\hline
1815    &  10.    &  1.9          &  94.4\%    & 93.0\% \\ 
\hline
1816    &  14.    &  2.5          &  93.4\%    & 91.0\% \\
\hline
1817    &  19     &  3.2          &  92.2\%    & 89.3\% \\
\hline
\end{tabular}}
\end{center}
\label{tab:veto-efficiency}   
\end{table}
An analysis of the experimental data with and without veto detectors
showed that the deflection of the electrons
by the magnet provided a sufficiently clean photon event sample.
As a result the veto hodoscopes were switched off during most high 
luminosity data-taking runs, although they proved important in analysis 
of low luminosity runs and in understanding various aspects of the experiment.
\section{High Voltage System}
\label{sec:HVS}
\begin{figure} [ht]
  \begin{center}
    \includegraphics[angle=0,width=\linewidth]{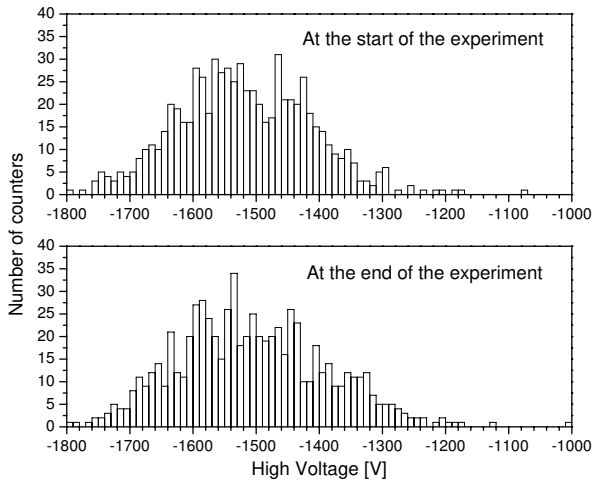}
    \caption{The HV settings for the calorimeter PMTs.}
    \label{fig:HV1}
  \end{center}
\end{figure}

Each PMT high-voltage supply was individually monitored and 
controlled by the High Voltage System (HVS). 
The HVS  consists of six power supply crates of LeCroy type 1458 
with high-voltage modules of type 1461N, a cable system, and 
a set of software programs.
The latter allows to control, monitor, download and save 
the high-voltage settings and is described below in more detail.
Automatic HV monitoring provides an alarm feature with 
a verbal announcement and a flashing signal on the terminal. 
The controls are implemented over an Ethernet network using TCP/IP protocol.
A Graphical User Interface (GUI) running on a Linux PC provides access to 
all features of the LeCroy system, loading the settings and saving
them in a file.  
A sample distribution of the HV settings 
is shown in Fig.~\ref{fig:HV1}.

The connections between the outputs of the high-voltage modules and the PMT 
dividers were arranged using 100~m long multi-wire cables.
The transition from the individual HV supply outputs to a multi-wire 
cable and back to the individual PMT was arranged via high-voltage 
distribution boxes that
are located inside the DAQ area and front-end patch panels 
outside the PMT box.
These boxes have input connectors for individual channels on one
side and two high-voltage multi-pin connectors
(27 pins from FISCHER part number D107 A051-27) on the other.
High-voltage distribution boxes were mounted on the side of the 
calorimeter stand and on the electronics rack.

\section{Data Acquisition System}
\label{sec:DAQ}
Since the calorimeter was intended to be used in Hall A at JLab together with 
the standard Hall A detector devices, the Data Acquisition System of the 
calorimeter is part of the standard Hall A DAQ system. 
The latter uses {\bf CODA} (CEBAF On-line Data Acquisition system) 
\cite{CODA} developed by the JLab data-acquisition group.
\begin{figure} [ht]
  \begin{center}
    \includegraphics[angle=0,width=\linewidth]{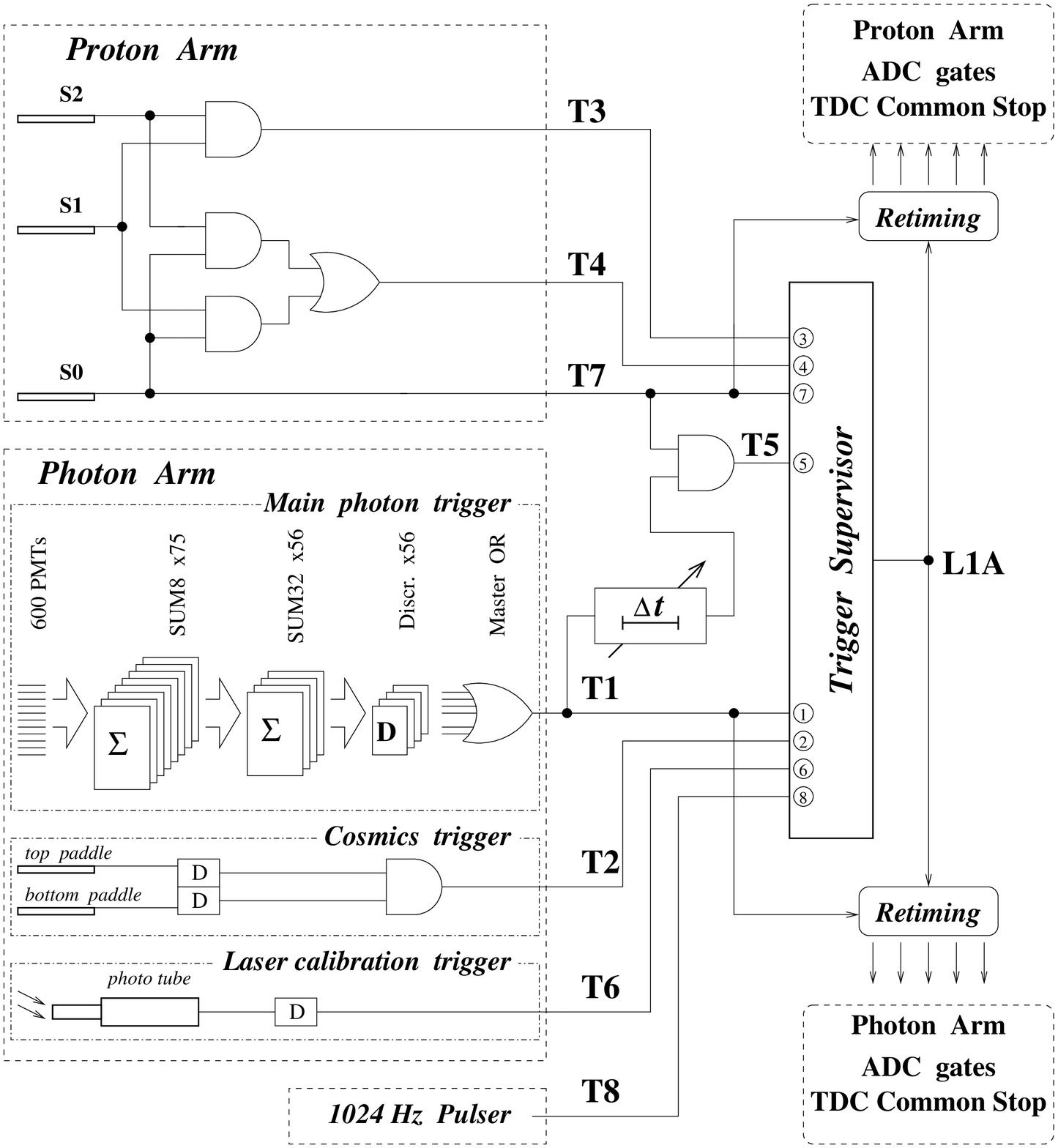}
     \caption{Schematic diagram of the DAQ trigger logic.}
    \label{fig:Trigger}
  \end{center}
\end{figure}
The calorimeter DAQ includes one Fastbus crate with a single-board VME 
computer installed using a VME-Fastbus interface and 
a trigger supervisor module \cite{TS}, which synchronizes 
the read-out of all the information in a given event. 
The most important software components are a Read-Out Controller
(ROC), which runs on the VME computer under the VxWorks OS, and 
an Event Builder and Event Recorder which both run on a Linux workstation. 
For a detailed description of the design and operation of the Hall A DAQ 
system see \cite{al04} and references therein.

All 704 PMT signals and 56 {\em SUM-32} signals are digitized by 
LeCroy 1881M FastBus ADC modules. 
The 56 {\em SUM-32} discriminator pulses are also
read-out by scalers and LeCroy 1877 FastBus TDCs.
During the RCS experiment the calorimeter
was operating in conjunction with one of the High Resolution Spectrometers
(HRS), which belong to the standard Hall A detector equipment \cite{al04}.
The Hall A Data Acquisition System is able to accumulate data 
involving several event types simultaneously. 
In the RCS experiment there were 8 types of trigger
signals and corresponding event types.
Trigger signals from the HRS are generated by three scintillator planes:
{\bf S0}, {\bf S1} and {\bf S2} (see Fig.~8 in \cite{al04}). 
In the standard configuration the main single arm trigger in 
the spectrometer is formed by a coincidence of signals from S1 and S2. 
An alternative trigger, logically described by
({\bf S0} AND {\bf S1}) OR ({\bf S0} AND {\bf S2}), is 
used to measure the trigger efficiency. 
In the RCS experiment one more proton arm trigger was used, defined as being 
a single hit in the {\bf S0} plane. 
As this is the fastest signal produced in the proton arm, it was better suited 
to form a fast coincidence trigger with the photon calorimeter.

The logic of the Photon Arm singles trigger was described in 
detail in Section~\ref{sec:electronics}. 
Besides this singles trigger there are 
two auxiliary triggers that serve to monitor the calorimeter blocks
and electronics. The first is a photon arm cosmics trigger, which was 
defined by a coincidence between signals from two plastic scintillator 
paddles, placed on top and under the bottom of the calorimeter. 
The other trigger is the light-calibration (laser) trigger which was 
used for gain monitoring purposes.

The two-arm coincidence trigger is formed by a time overlap of 
the main calorimeter trigger and the signal from the {\bf S0} scintillator
plane in the HRS. 
The width of the proton trigger pulse is set to 100 ns, while 
the photon trigger pulse, which is delayed in a programmable delay line, 
is set to 10~ns.
As a result, the coincidence events are synchronized with the photon 
trigger, and a correct timing relation between trigger signals from 
two arms is maintained for all 25 kinematic configurations of 
the RCS experiment.
Finally, a 1024 Hz puls generator signal forms a pulser trigger, which
was used to measure the dead time of the electronics.

All 8 trigger signals are sent to the Trigger Supervisor module which
starts the DAQ readout. Most inputs of the Trigger Supervisor can
be individually pre-scaled. Triggers which are accepted by the DAQ are
then re-timed with the scintillators of a corresponding arm to make 
gates for ADCs and TDCs. This re-timing removes trigger time jitter 
and ensures the timing is independent of the trigger type.
Table~\ref{trig_table} includes information on the trigger
and event types used in the RCS experiment and shows typical pre-scale
factors used during the data-taking.
A schematic diagram of the overall RCS experiment DAQ trigger logic 
is shown in Fig.\ref{fig:Trigger}.

\begin{table}[hbt]
\caption{\label{trig_table}
A list of triggers used in the RCS experiment. Typical pre-scale
factors which were set during a data-taking run (run \#1819) are
shown.}
\begin{center}
\begin{tabular}{|c|p{0.5\linewidth}|c|}
\hline
Trigger & Trigger Description & pre-scale \\
  ID    &                     &  factor   \\ \hline
T1 & Photon arm singles trigger & 100,000 \\
T2 & Photon arm cosmics trigger & 100,000 \\
T3 & Main Proton arm trigger: \mbox{(S1 {\it AND} S2)} & 1 \\
T4 & Additional Proton arm trigger: 
      \mbox{({\bf S0} {\it AND} {\bf S1}) OR ({\bf S0} {\it AND} {\bf S2})} & 10 \\
T5 & Coincidence trigger & 1 \\
T6 & Calorimeter light-calibration trigger & 1 \\
T7 & Signal from the HRS {\bf S0} scintillator plane & 65,000 \\
T8 & 1024 Hz pulser trigger & 1,024 \\
\hline 
\end{tabular}
\end{center}
\end{table}

\section{Calorimeter Performance}
The calorimeter used in the RCS experiment had three related purposes.
The first purpose is to 
provide a coincidence trigger signal for operation of the DAQ.
\begin{figure} [ht]
  \begin{center}
    \includegraphics[angle=0,width= 0.9 \linewidth]{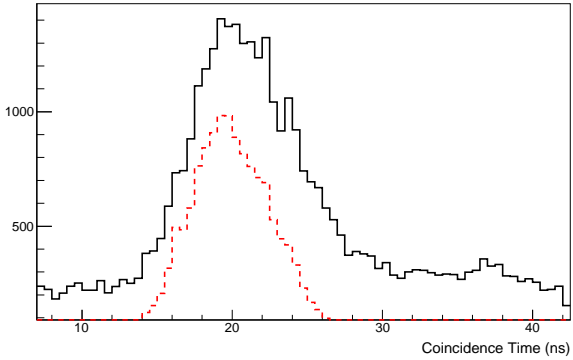}
    \caption{The time of the calorimeter trigger relative to 
the recoil proton trigger for a production run in kinematic $3E$ 
at maximum luminosity (detected $E_\gamma = 1.31~GeV$). 
The solid curve shows all events, while the dashed curve shows events with a cut
on energy in the most energetic cluster $>1.0$~GeV.}    
\label{fig:coin-time}
  \end{center}
\end{figure}
Fig.~\ref{fig:coin-time} shows the coincidence time distribution, where
one can see a clear relation between energy threshold and 
time resolution. The observed resolution of around 8 ns (FWHM) was sufficient to identify 
cleanly coincidence events over the background, which meant that 
no off-line corrections were applied for variation of the average time of 
individual $SUM-32$ summing modules. The second purpose is determination 
of the energy of the scattered photon/electron 
to within an accuracy of a few percent, while the third is reasonably accurate 
reconstruction of the photon/electron hit 
coordinates in order that kinematic correlation cuts between the scattered 
photon/electron and the recoil proton can be made.
The off-line analysis procedure and the observed position and energy resolutions 
are presented and discussed in the following two sections.

\subsection{Shower Reconstruction Analysis and Position Resolution}
The off-line shower reconstruction involves a search for clusters 
and can be characterized by the following definitions:
\begin{enumerate}
\item a cluster is a group of adjacent blocks; 
\item a cluster occupies 9 ($3 \times 3$) blocks of the calorimeter; 
\item the distribution of the shower energy deposition over the cluster blocks
  (the so-called shower profile) satisfies the following conditions: 
\begin{enumerate}
\item the maximum energy deposition is in the central block; 
\item the energy deposition in the corner blocks is less than that in each of two 
      neighboring blocks;
\item around 50\% of the total shower energy must be deposited in the central row 
(and column) of the cluster. 
\end{enumerate}
\end{enumerate}
For an example in which the shower center is in the middle of the central block, 
around 84\% of the total shower energy is in the central block, about 
14\% is in the four neighboring blocks, and the remaining 2\% is in the corner blocks.
Even at the largest luminosity used in the RCS experiment the probability 
of observing two clusters with energies above 50\% of the elastic value 
was less than 10\%, so for the 704 block hodoscope a two-cluster 
overlap was very unlikely.

The shower energy reconstruction requires both hardware and software 
calibration of the calorimeter channels. 
On the hardware side, equalization of the counter gains was initially done 
with cosmic muons,
which produce 20~MeV energy equivalent light output per 4~cm path 
(muon trajectories perpendicular to the long axis of the lead-glass blocks).
The calibration was done by selecting cosmic events for which the signals 
in both counters above and below a given counter were large.
The final adjustment of each counter's gain was done by using calibration
with elastic e-p events. This calibration provided PMT gain values which 
were on average different from the initial cosmic set by 20\%

The purpose of the software calibration is to define the coefficients for 
transformation of the ADC amplitudes to energy deposition for 
each calorimeter module. 
These calibration coefficients are obtained from elastic e-p data 
by minimizing the function:
\begin{eqnarray*}
  \chi^2 = \sum_{n=1}^N \Big[ 
  \sum_{i \in M^n} C_i \cdot (A_i^n-P_i) - E_e^n \Big]^2
\end{eqnarray*}
where: \\
\hspace*{.50cm} $n=1 \div N$ --- number of the selected calibration event;\\
\hspace*{.50cm} $i$ --- number of the block, included in the cluster;\\
\hspace*{.50cm} $M^n$ --- set of the blocks' numbers, in the cluster;\\
\hspace*{.50cm} $A_i^n$ --- amplitude into the $i$-th block;\\
\hspace*{.50cm} $P_i$ --- pedestal of the $i$-th block;\\
\hspace*{.50cm} $E_e^n$ --- known energy of electron;\\
\hspace*{.50cm} $C_i$ --- calibration coefficients, which need to be fitted.\\
The scattered electron energy $E_e^n$ is calculated by using the energy of the 
primary electron beam and the scattered electron angle. 
A cut on the proton momentum-angle correlation is used to select 
clean elastic events.
   
Following calculation of the calibration coefficients, the total energy 
deposition $E$, as well as the $X$ and $Y$ coordinates of the shower center 
of gravity are calculated by the formulae: 
\begin{eqnarray*}
  E=\sum_{i \in M}E_i\ ,\ \ \ \ 
  X=\sum_{i \in M}E_i \cdot X_i/E\ ,\ \ \ \ 
  Y=\sum_{i \in M}E_i \cdot Y_i/E\ 
\end{eqnarray*}
where $M$ is the set of blocks numbers which make up the cluster, 
$E_i$ is the energy deposition in the $i$-th block, and 
$X_i$ and $Y_i$ are the coordinates of the $i$-th block center.
The coordinates calculated by this simple center of gravity method 
are then used for a more accurate determination of the incident hit position. 
This second iteration was developed during the second test run \cite{ch98}, 
in which a two-layer MWPC was constructed 
and positioned directly in front of the calorimeter.
This chamber had 128 sensitive wires in both X and Y directions, with a
wire spacing of 2~mm and a position resolution of 1~mm.
In this more refined procedure, the coordinate $x_o$ of the shower center of gravity 
inside the cell (relative to the cell's low boundary) is used. 
An estimate of the coordinate $x_e$ can be determined from a polynomial in this
coordinate ($P(x_o)$): 
\begin{eqnarray*}
  x_e = P(x_o) = a_1 \cdot x_o + a_3 \cdot x^3_o + a_5 \cdot x^5_o + 
                                 a_7 \cdot x^7_o + a_9 \cdot x^9_o
\end{eqnarray*}
For symmetry reasons, only odd degrees of the polynomial are used.
The coefficients $a_n$ are calculated by minimizing the functional:
\begin{eqnarray*}
  \chi^2 = \sum_{i=1}^N \Big[ P(a_n,x^i_o) - x^i_t \Big]^2.
\end{eqnarray*}
where: \\
\hspace*{.50cm} $i=1 \div N$ --- number of event;\\
\hspace*{.50cm} $x^i_o$ --- coordinate of the shower center of gravity inside 
                the cell;\\
\hspace*{.50cm} $x^i_t$ --- coordinate of the track (MWPC) 
on the calorimeter plane;\\
\hspace*{.50cm} $a_n$ --- coordinate transformation 
coefficients to be fitted. \\

\begin{figure} [htb]
  \begin{center}
    \includegraphics[angle=0,width=0.9\linewidth]{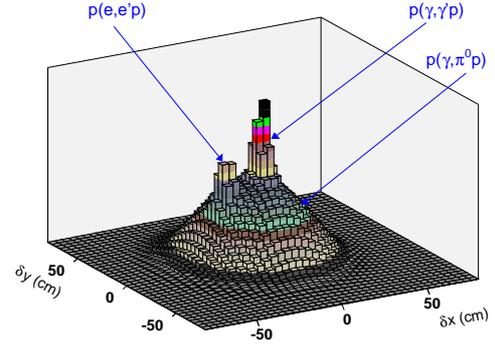}
    \caption{The scatter plot of $p-\gamma (e)$ events in the 
plane of the calorimeter front face.}
    \label{fig:dydx}
  \end{center}
\end{figure}

The resulting resolution obtained from such a fitting procedure was 
found to be around 5.5~mm for a scattered electron energy of 2.3~GeV. 
For the case of production data, where the MWPC was not used, 
Fig.~\ref{fig:dydx} shows a scatter plot of events on the front face 
of the calorimeter. The parameter plotted is 
the differences between the observed hit coordinates in 
the calorimeter and the coordinates calculated from the proton parameters 
and an assumed two-body kinematic correlation. The dominant contribution to 
the widths of the RCS and e-p peaks that can be seen in this figure is from 
the angular resolution of the detected proton, which is itself dominated by
multiple scattering. As the calorimeter distance varied during the experiment 
between 5.5~m and 20~m, the contribution to the combined angular resolution 
from the calorimeter position resolution of a few millimeters was minimal. 

\subsection{Trigger Rate and Energy Resolution}
\label{sec:Rate}
%
At high luminosity, when a reduction of the accidental coincidences
in the raw trigger rate is very important, the trigger threshold should be set
as close to the signal amplitude for elastic RCS photons as practical.
However, the actual value of the threshold for an individual event has a
significant uncertainty due to pile-up of the low-amplitude signals,
fluctuations of the signal shape (mainly due to summing of the
signals from the PMTs with different HV and transit time),
and inequality of the gain in the individual counters. Too high a threshold,
therefore, can lead to a loss in detection efficiency.

\begin{figure}[ht]
 \begin{center}
     \includegraphics[angle=0,width=\linewidth]{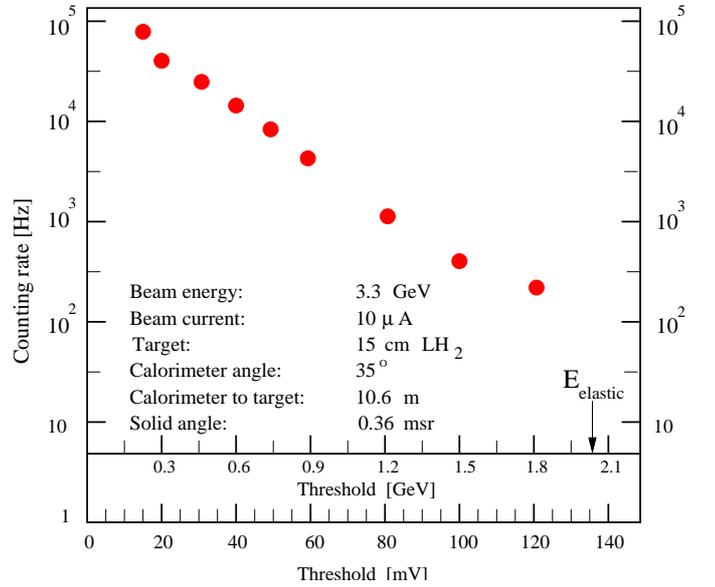}
      \caption{Calorimeter trigger rate vs threshold level.}
    \label{fig:calo-rate}
 \end{center}
\end{figure}

The counting rate of the calorimeter trigger, $f$, which defines a practical 
level of operational luminosity has an exponential dependence on the threshold,
as can be seen in Fig.~\ref{fig:calo-rate}.
It can be described by a function of $E_{thr}$:
 \begin{eqnarray*}
f \,=\, A \times \exp( -B \times E_{thr} / E_{max} ),
 \end{eqnarray*}
where $E_{max}$ is the maximum energy of an elastically scattered 
photon/electron for a given scattering angle, $A$ an angle-dependent 
constant, and $B$ a universal constant $\approx 9 \pm 1 $.
The angular variation of the constant $A$, after normalization to a fixed 
luminosity and the calorimeter solid angle, is less than 
a factor of 2 for the RCS kinematics. The threshold for all kinematics was
chosen to be around half of the elastic energy, thereby balancing the need
for a low trigger rate without affecting the detection efficiency. 

\begin{figure} [htb]
  \begin{center}
    \includegraphics[angle=0,width=0.9\linewidth]{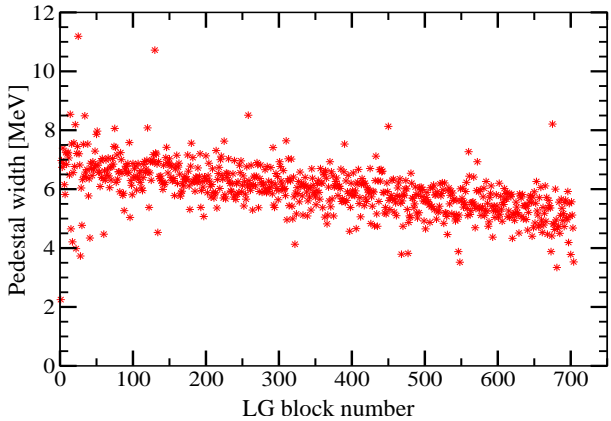}
    \caption{The width of the ADC pedestals for the calorimeter in a typical run.
The observed reduction of the width vs the block number reflects the lower 
background at larger detector angle with respect to the beam direction.}
    \label{fig:pedestals}
  \end{center}
\end{figure}

In order to ensure proper operation and to monitor the performance of each counter 
the widths of the ADC pedestals were used (see Fig.~\ref{fig:pedestals}).
One can see that these widths vary slightly with block number, which reflects 
the position of the block in the calorimeter and its angle with respect to 
the beam direction. This pedestal width also allows for an estimate of the 
contribution of the background induced base-line fluctuations to the overall 
energy resolution. For the example shown in Fig.~\ref{fig:pedestals}
the width of 6~MeV per block leads to energy spectrum noise of about 20~MeV 
because a 9-block cluster is used in the off-line analysis.

The energy resolution of the calorimeter was measured by using elastic 
e-p scattering. Such data were collected many times during the experiment for 
kinematic checks and calorimeter gain calibration. Table~\ref{tab:resolution} 
presents the observed resolution and the corresponding ADC pedestal widths 
over the course of the experiment. For completeness, the pedestal widths for 
cosmic and production data are also included.
%
\begin{sidewaystable*}
\begin{center}
\caption{Pedestal widths and calorimeter energy resolution at different stages
of the RCS experiment for cosmic (c), electron (e) and production ($\gamma$) runs in 
order of increasing effective luminosity.
\vspace*{10pt}}
{\footnotesize
\begin{tabular}{|l|c|c|c|c|c|c|c|c|}
\hline
Runs    &$\cal L$ $_{eff}$& Beam Current & Accumulated & ~Detected~$E_{e/\gamma}$~&~$\sigma_{_E}/E$~&~$\sigma_{_E}/E$~at E$_\gamma$=1 GeV~ &~$\Theta_{cal}$~&  ~$\sigma_{ped}$~   \\ 
        & ($10^{38}$ cm$^{-2}$/s) &($\mu$A) &~ Beam Charge (C)~ & (GeV)         &     (\%)  &  (\%)  &~(degrees)~&~(MeV)~\\
\hline
1517 (c)      & -       & -       &      -  & -    & -   & -  &  -   &  1.5     \\ 
\hline
1811 (e)      &  0.1   &  2.5    &  2.4    & 2.78  & 4.2 & 7.0  &  30  &  1.7   \\ 
\hline
1488 (e)      &  0.2   &  5    &  0.5    & 1.32    & 4.9 & 5.5 &  46  &  1.75   \\ 
\hline
2125 (e)      &  1.0   &  25    &  6.6    & 2.83   & 4.9 & 8.2  &  34  &  2.6   \\ 
\hline
2593 (e)      &  1.5   &  38    & 14.9    & 1.32   & 9.9 & 11.3  &  57  & 2.0   \\
\hline
1930 (e)      &  1.6   &  40    &  4.4    & 3.39   & 4.2 & 7.7  &  22  &  3.7   \\ 
\hline
1938 ($\gamma$) &  1.8   &  15  &  4.5      & 3.23   & - & -  &  22  &  4.1   \\ 
\hline
2170 ($\gamma$) &  2.4   &  20  &   6.8     & 2.72   & - & -  &  34  &  4.0   \\ 
\hline
1852 ($\gamma$) &  4.2   &   35  &   3.0     & 1.63   & - & -  &  50  &  5.0   \\ 
\hline
\end{tabular}}
\label{tab:resolution}   
\end{center}
\end{sidewaystable*}
%
At high luminosity the energy resolution degrades due to fluctuations 
of the base line (pedestal width) and the inclusion of more accidental hits 
during the ADC gate period.
However, for the 9-block cluster size used in the data analysis 
the contribution of the base line fluctuations to the energy resolution 
is just 1-2\%.
The measured widths of ADC pedestals confirmed the results of Monte Carlo
simulations and test runs that the radiation background is three times higher 
with the 6\% Cu radiator upstream of the target than without it.

The resolution obtained from e-p calibration runs was corrected 
for the drift of the gains so it could be attributed directly 
to the effect of lead glass radiation damage.
It degraded over the course of 
the experiment from 5.5\% (for a 1~GeV photon energy) at the start to
larger than 10\% by the end.
It was estimated that this corresponds to a 
final accumulated radiation dose of
about 3-10~kRad, which is in agreement with the known level of 
radiation hardness of the TF-1 lead glass \cite{in83}. 
This observed radiation dose corresponds to a 500 hour experiment 
with a 15~cm LH2 target and 50~$\mu$A beam. 

\subsection{Annealing of the radiation damage}

\begin{figure}[ht] 
 \begin{center}
   \includegraphics[angle=0, width= 0.9 \linewidth]{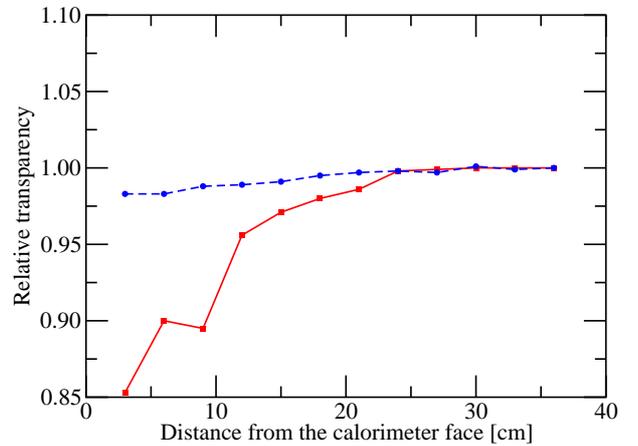}
   \caption{The blue light attenuation in 4~cm of lead-glass 
vs distance from the front face of calorimeter measured before (solid) 
and after (dashed) UV irradiation.}
   \label{fig:rad_glass}
 \end{center}
\end{figure}

The front face of the calorimeter during the experiment was protected
by plastic material with an effective thickness of 10~g/cm$^2$.  
For the majority of the time the calorimeter was located at a distance 
of 5-8~m and an angle of 40-50$^\circ$ with respect to the electron 
beam direction.
The transparency of 20 lead-glass blocks was measured after the 
experiment, the results of which are shown in Fig.~\ref{fig:rad_glass}. 
This plot shows the relative transmission through 4~cm of glass in 
the direction transverse to the block length at different locations. 
The values were normalized to the transmission through similar 
lead-glass blocks which were not used in the experiment. 
The transmission measurement was done with a blue LED 
($\lambda_{max}$ of 430~nm) and a Hamamatsu photo-diode (1226-44).

A UV technique was developed and used in order to cure radiation damage.
The UV light was produced by a 10~kW total power 55-inch long 
lamp\footnote{Type A94551FCB manufactured by American Ultraviolet, 
Lebanon, IN 46052, USA}, which was installed vertically at a 
distance of 45 inches from the calorimeter face and a quartz plate 
(C55QUARTZ) was used as an infrared filter.
The intensity of the UV light at the face of the lead-glass blocks
was found to be 75 mW/cm$^2$ by using a UVX digital radiometer\footnote{
Manufactured by UVP, Inc., Upland, CA 91786, USA}.
In situ UV irradiation without disassembly of the lead-glass stack 
was performed over an 18 hour period.
All PMTs were removed before irradiation to ensure the
safety of the photo-cathode. The resultant improvement in transparency
can be seen in Fig.~\ref{fig:rad_glass}.
An alternative but equally effective method to restore the 
lead-glass transparency, which involved heating of the lead-glass blocks 
to 250$^\circ$C for several hours, was also tested. The net effect of heating 
on the transparency of the lead-glass was similar to the UV curing results.

In summary, operation of the calorimeter at high luminosity, particularly when 
the electron beam was incident on the bremsstrahlung radiator, led to a degradation 
in energy resolution due to fluctuations in the base-line and a higher accidental rate 
within the ADC gate period. For typical clusters this effect was found to be around a 
percent or two. By far the largest contributor to the observed degradation in resolution 
was radiation damage sustained by the lead-glass blocks, which led to the resolution 
being a factor of two larger at the end of the experiment. The resulting estimates 
of the total accumulated dose were consistent with expectations for this type of 
lead-glass. Finally, it was found that both UV curing and heating of the lead-glass 
were successful in annealing this damage.

\section{Summary}
\label{sec:summary}

The design of a segmented electromagnetic calorimeter which was used
in the JLab RCS experiment has been described. The performance of the 
calorimeter in an unprecedented high luminosity, high background
environment has been discussed. Good energy and position 
resolution enabled a successful measurement of the RCS process over
a wide range of kinematics.

\section{Acknowledgments}
\label{sec:thanks}

We acknowledge the RCS collaborators who helped to operate the detector and the
JLab technical staff for providing outstanding support, and specially
D.~Hayes, T.~Hartlove, T.~Hunyady, and S.~Mayilyan for help in the construction
of the lead-glass modules.
We appreciate S.~Corneliussen's careful reading of the manuscript
and his valuable suggestions.
This work was supported in part by the National Science Foundation in grants 
for the University of Illinois University 
and by DOE contract DE-AC05-84ER40150 under which the Southeastern Universities 
Research Association (SURA) operates the Thomas Jefferson National Accelerator 
Facility for the United States Department of Energy.

\end{document}